\documentclass[twocolumn,twoside]{revtex4}
\usepackage{graphicx}
\usepackage{fancyhdr}
\usepackage{makecell}
\begin{document}

\title{$\Lambda_{c}$ decays at BESIII}

%

\author{Y.~Gao\\
	on behalf of the BESIII}
\affiliation{University of Science and Technology of China,~Hefei 230026,~People's Republic of China\\
State Key Laboratory of Particle Detection and Electronics,~Beijing 100049,~Hefei 230026,~People's Republic of China}
%

%

\begin{abstract}
BESIII has made great progress in taking data which is the largest data samples near the $\Lambda_{c}^{+}\bar{\Lambda}_{c}^{-}$ threshold. We have measured the branching fraction of $\Lambda_{c}^{+}\to n\pi^{+}$ to be $(6.6\pm1.2_{\mathrm{stat.}}\pm0.4_{\mathrm{syst.}})\times10^{-4}$ with the statistical significance of $7.3~\sigma$ firstly using 3.9$\mathrm{fb}^{-1}$ $e^{+}e^{-}$ collision collected with BESIII detector at six center-of-mass energies between 4.612 and 4.699 GeV. With the result of the branching fraction of $\Lambda_{c}^{+}\to p\pi^{0}$ from Belle, the ratio of the branching fractions between $\Lambda_{c}^{+}\to n\pi^{+}$ and $\Lambda_{c}^{+}\to p \pi^{0}$ is measured to be larger than 7.2 at 90\% confidence level. The branching fractions of $\Lambda_{c}^{+}\to\Lambda\pi^{+}$ and $\Lambda_{c}^{+}\to\Sigma^{0}\pi^{+}$ are measured to be $(1.31\pm0.08_{\mathrm{stat.}}\pm0.05_{\mathrm{syst.}})\times10^{-2}$ and $(1.22\pm0.08_{\mathrm{stat.}}\pm0.07_{\mathrm{syst.}})\times10^{-2}$, respectively, which are consistent with previous BESIII results. More results of $\Lambda_{c}$ decays will be published with better precision this year using 6.4$\mathrm{fb}^{-1}$ $e^{+}e^{-}$ collision data samples between 4.600 and 4.946 GeV.

\end{abstract}

\maketitle

\thispagestyle{fancy}


\section{Introduction}

$\Lambda_{c}^{+}$ is the ground states of singly charmed baryon and most of the charmed baryons will eventually decay to $\Lambda_{c}^{+}$ but the knowledge of $\Lambda_{c}^{+}$ decays is still very limited in comparison with charmed mesons. Firstly, only a single semileptonic (SL) decay mode $\Lambda_{c}^{+} \to \Lambda l^{+} \nu_{l}$ has been observed. Secondly, the doubly Cabibbo-suppressed (DCS) modes have not been systematically studied. Thirdly, the known exclusive decays of $\Lambda_{c}^{+}$ account for only about 60\% of the total branching fraction. In particular, information about the decays that involve a neutron is minimal, considering that such decays should account for nearly half of the $\Lambda_{c}$ decay rate. Fourth, the precision of singly Cabibbo-suppressed (SCS) modes is very limit and some two-body SCS modes still need to be searched. Improved knowledge of the $\Lambda_{c}^{+}$ decays is essential for the studies of the whole charmed baryon family~\cite{WhitePaper}. 

Several different phenomenological models, e.g. current algebra~\cite{Current1,Current2} and SU(3) flavor symmetry~\cite{SU(3)1,SU(3)2,SU(3)3}, have been employed as tools to reveal the dynamics of charmed baryon decays. The hadronic decay amplitudes of $\Lambda_{c}^{+}$ consist of factorizable and nonfactorizable components. The non-factorizable contributions are important in these decays which can be constrained by measurements. So it is important to get comprehensive and precise experimental inputs to understand the validity of different phenomenological models. 

The two-body SCS decay $\Lambda_{c}^{+}\to n\pi^{+}$, together with $\Lambda_{c}^{+}\to p\pi^{0}$ are of great interest and have been studied extensively in the context of various phenomenological models~\cite{SU(3)1,SU(3)2,Uppal1994,Chen2003,Cheng2018,Geng2018,Geng2019,Geng2020,Zou2020,Zhao2020}. They predict quite different decay rates for $\Lambda_{c}^{+}\to p\pi^{0}$ and $\Lambda_{c}^{+}\to n\pi^{+}$ and distinguishing between these models with experimental results is highly desirable. This ratio is predicted to be 2 by the SU(3) flavor symmetry model~\cite{SU(3)1,SU(3)2,Geng2018}, 4.5 or 8.0 by the constituent quark model~\cite{Uppal1994}, 3.5 by a dynamical calculation based on pole model and current-algebra~\cite{Cheng2018}, 4.7 by the SU(3) flavor symmetry including the contributions from $\mathcal{O}(\overline{15})$~\cite{Geng2019}, and 9.6 by the topological-diagram approach~\cite{Zhao2020}.

To date, the branching fractions of only a few SCS decay modes have been measured, and all with limited precision~\cite{PDG2020}, and those involving a neutron in the final state have never been measured. So it is also important to search the decay $\Lambda_{c}^{+} \to n \pi^{+}$ and other SCS decay modes of $\Lambda_{c}^{+}$.

\section{The BESIII detector}
The BESIII detector~\cite{BESIIIDector2010} records symmetric $e^{+}e^{-}$ collisions provided by the BEPCII storage ring~\cite{BESIIIDector2016}, which operates in the center-of-mass energy range from 2.0 GeV to 4.946 GeV. BESIII has collected large data samples in this energy region~\cite{BESIIIDector2020}. The cylindrical core of the BESIII detector covers 93\% of the full solid angle and consists of a helium-based multilayer drift chamber (MDC), a plastic scintillator time-of-flight system (TOF), and a CsI(Tl) electromagnetic calorimeter (EMC), which are all enclosed in a superconducting solenoidal magnet providing a 1.0 T magnetic field. The solenoid is supported by an octagonal flux-return yoke with resistive plate counter muon identification modules interleaved with steel. The charged-particle momentum resolution at 1 $\mathrm{GeV}/c$ is 0.5\%, and the $dE/dx$ resolution is 6\% for electrons from Bhabha scattering. The EMC measures photon energies with a resolution of 2.5\% (5\%) at 1 GeV in the barrel (end cap) region. The time resolution in the TOF barrel region is 68 ps, while that in the end cap region is 60 ps~\cite{BESIIIDector20202}.

\section{Recent Results}

\subsection{Observation of the Singly Cabibbo-Suppressed Decay $\Lambda_{c}^{+}\to n \pi^{+}$}

The first observation of the SCS decay $\Lambda_{c}^{+}\to n \pi^{+}$ is reported using $3.9\mathrm{fb}^{-1}$ $e^{+}e^{-}$ collision data collected with the BESIII detector at six center-of-mass energies between 4.612 and 4.699 GeV. The integrated luminosities of the data samples at 4.612, 4.628, 4.641, 4.682, and 4.699 GeV are 103.5, 519.9, 548.2, 527.6, 1664.3, and 534.4 $\mathrm{pb}^{-1}$~\cite{Lum}, respectively. Charge-conjugate modes are implicitly included.

Simulated samples are produced with a GEANT4-based Monte Carlo (MC) package~\cite{Geant4}. The detector response are used to determine detection efficiencies and to estimate backgrounds. A detailed description of the MC of this work can be found in~\cite{npi}.

\begin{figure}[!htp]
	\begin{center}
		\includegraphics[width=80mm]{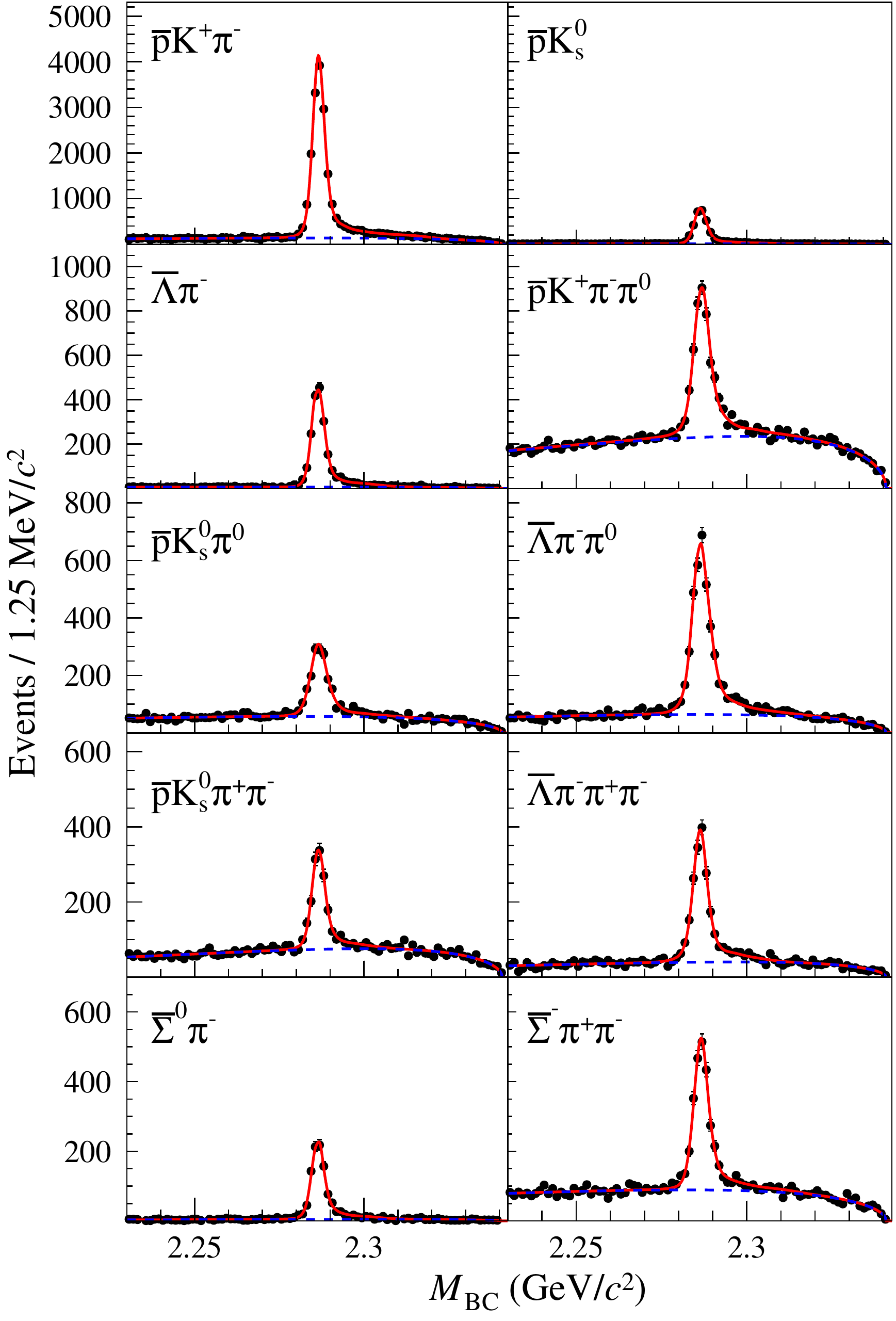}
	\end{center}
	\caption{
		The $M_\mathrm{BC}$ distributions of the ST modes for data sample at 
		$\sqrt{s}=4.682~\mathrm{GeV}$. 
		The points with error bars represent data. The (red) solid curves indicate the fit results and the (blue) dashed curves describe
		the background shapes.}
	\label{fig:single-tag-468}
\end{figure}

A double-tag (DT) approach~\cite{DTMethod} is implemented to search for $\Lambda_{c}^{+}\to n \pi^{+}$. A data sample of $\bar{\Lambda}_{c}^{-}$ baryons, referred to as the single-tag (ST) sample, is reconstructed with ten exclusive hadronic decay modes shown in FIG.~\ref{fig:single-tag-468}, where the intermediate particles $K_{S}^{0}$, $\bar{\Lambda}$, $\bar{\Sigma}^{0}$, $\bar{\Sigma}^{-}$, and $\pi^{0}$ are reconstructed with the decays $K_{S}^{0}\to\pi^{+}\pi^{-}$, $\bar{\Lambda}\to\bar{p}\pi^{+}$, $\bar{\Sigma}^{0}\to \gamma\bar{\Lambda}$, $\bar{\Sigma}^{-}\to\bar{p}\pi^{0}$, and $\pi^{0}\to \gamma\gamma$, respectively. Those events in which the signal decay $\Lambda_{c}^{+}\to n\pi^{+}$ is reconstructed in the system recoiling against the $\bar{\Lambda}_{c}^{-}$ candidates of the ST sample are denoted as DT candidates.

A detailed description of the selection of the ST sample is shown in~\cite{npi}. The decay $\Lambda_{c}^{+}\to n \pi^{+}$ is searched for among the remaining tracks recoiling against the ST $\bar{\Lambda}_{c}^{-}$ candidates. Only one tight charged track is allowed, which is then assigned to be the $\pi^{+}$ from the signal decay. To suppress contamination from long-lifetime particles in the final state, the candidate events are further required to be without any loose tracks.

The $\Lambda_{c}^{+}$ decay branching fractions ($\mathcal{B}$) are determined as 
\begin{equation} \label{eq:br}
	\mathcal{B}=\frac{N_{\mathrm{obs}}} {\sum_{ij} N_{ij}^{\mathrm{ST}}\cdot (\epsilon_{ij}^{\mathrm{DT}}/\epsilon_{ij}^{\mathrm{ST}}) },
\end{equation}
where the subscripts $i$ and $j$ represent the ST modes and the data samples at different c.m.~energies, respectively.
The parameters $N_{ij}^{\mathrm{ST}}$, $\epsilon_{ij}^{\mathrm{ST}}$ and $\epsilon_{ij}^{\mathrm{DT}}$ are the ST yields, ST and DT efficiencies, respectively.
The detection efficiencies $\epsilon_{ij}^{\mathrm{ST}}$ and $\epsilon_{ij}^{\mathrm{DT}}$ are estimated from MC samples.

The ST $\bar{\Lambda}_{c}^{-}$ candidates are identified using the variables of beam-constrained invariant mass $M_\mathrm{BC} = \sqrt{E_{\mathrm{beam}}^2/c^4 - | \vec{p}_{\bar{\Lambda}_{c}^{-}} |^2/c^2}$, where $E_{\mathrm{beam}}$ is the beam energy, $\vec{p}_{\bar{\Lambda}_{c}^{-}}$ is the momentum of the $\bar{\Lambda}_{c}^{-}$ candidate. The $M_{\mathrm{BC}}$ distributions of surviving candidates for the ten ST modes  are illustrated in FIG.~\ref{fig:single-tag-468} for the data sample at $\sqrt{s}=4.682~\mathrm{GeV}$, where clear $\bar{\Lambda}_{c}^{-}$ signals are observed in each sample. No peaking backgrounds are found with the investigation of the inclusive MC sample. The sum of the ST yields for all the six data samples is $90,692\pm359$, where the uncertainty is statistical.

To improve detection efficiency, the neutron is selected through  the recoiling mass ($M_{\rm rec}$) against the ST $\bar{\Lambda}_{c}^{-}$ and $\pi^+$:
\begin{equation} \label{eq:mrec}
	M^2_{\rm rec} = (E_{\mathrm{beam}} - E_{\pi^+})^2/c^4 - | \rho\cdot\vec{p}_{0} - \vec{p}_{\pi^+} |^2/c^2 ,
\end{equation}
where $E_{\pi^+}$ and $\vec{p}_{\pi^+}$ are the energy and momentum of the $\pi^+$ candidate,
$\rho = \sqrt{E_{\mathrm{beam}}^2/c^2 - m_{\Lambda_{c}^{+}}^2 c^2}$, and $\vec{p}_{0} = - \vec{p}_{\bar{\Lambda}_{c}^{-}}/|\vec{p}_{\bar{\Lambda}_{c}^{-}}|$ is the unit direction opposite to the ST $\bar{\Lambda}_{c}^{-}$. After imposing all selection conditions, the distribution of  $M_{\rm rec}$ of the accepted DT candidate events from the combined six data samples at different c.m.~energies is shown in FIG.~\ref{eq:mrec}, where a peak at the neutron mass is observed, representing the  $\Lambda_{c}^{+}\to n \pi^{+}$ signal. Additionally, there are two prominent structures peaking at the $\Lambda$ and $\Sigma^0$ mass regions, which represent the Cabibbo-favored (CF) processes $\Lambda_{c}^{+}\to \Lambda \pi^{+}$ and $\Lambda_{c}^{+}\to \Sigma^{0} \pi^{+}$, respectively. The potential backgrounds can be classified into two categories: those directly originating from continuum hadron production in the $e^{+}e^{-}$ annihilation (referred to as $q\bar{q}$ background hereafter) and those from $e^{+}e^{-} \to \Lambda_{c}^{+} \bar{\Lambda}_{c}^{-}$ events (referred to as  $\Lambda_{c}^{+} \bar{\Lambda}_{c}^{-}$ background hereafter), excluding contributions from $\Lambda_{c}^{+}\to n \pi^{+}$, $\Lambda\pi^+$, and $\Sigma^0\pi^+$ signals.

\begin{figure}[!htp]
	\begin{center}
		\includegraphics[width=80mm]{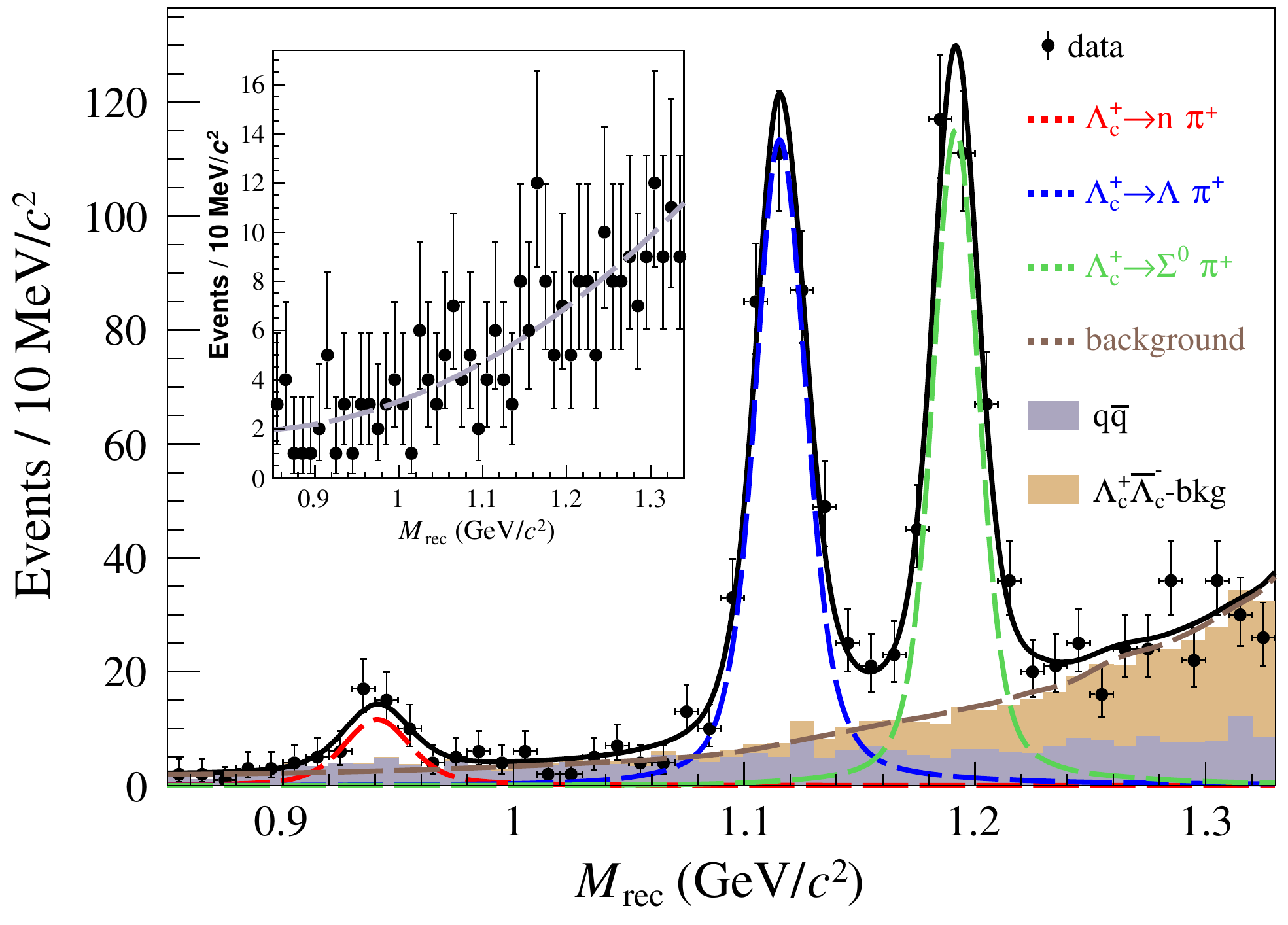}   
	\end{center}
	\caption{ The $M_{\mathrm{rec}}$ distribution of the accepted DT candidate events from the combined six data samples.
		The black points with error bars are data. The red, blue, and green dashed lines indicate the curves for the neutron, 
		$\Lambda$, and $\Sigma$ peaks, respectively. 
		The brown and gray shaded histograms for the two background components are from the inclusive MC sample, and the dark brown dashed line indicates the curve that describes the two background components from fitting.
		The black line is the sum over all the components in the fit. The inset shows the $M_{\mathrm{rec}}$ distribution in the ST $M_\mathrm{BC}$ sideband region, and the gray dashed line indicates the curve that describes the $q\bar{q}$ background.}
	\label{fig:fit}
\end{figure}

The fit distributions are depicted in FIG.~\ref{eq:mrec}, and correspond to signal yields of $50 \pm 9$,  $376 \pm 22$, and $343 \pm 22$ for the decays $\Lambda_{c}^{+}\to n \pi^{+}$, $\Lambda \pi^+$ and $\Sigma^0 \pi^+$, respectively, where the uncertainties are statistical and the statistical significance of $\Lambda_{c}^{+}\to n \pi^{+}$ is $7.3\sigma$.

The branching fraction of $\Lambda_{c}^{+}\to n \pi^{+}$ is measured to be $(6.6\pm1.2_{\rm stat}\pm0.4_{\rm syst})\times 10^{-4}$, this is a first-time measurement. Meanwhile, the branching fractions of the Cabibbo-favored decays $\Lambda_{c}^{+}\to \Lambda \pi^{+}$ and $\Lambda_{c}^{+}\to \Sigma^{0} \pi^{+}$ 
are measured to be $(1.31\pm0.08_{\rm stat}\pm0.05_{\rm syst})\times 10^{-2}$ and  $(1.22\pm0.08_{\rm stat}\pm0.07_{\rm syst})\times 10^{-2}$, respectively. Taking the upper limit of the branching fraction of $\Lambda_{c}^{+}\to p\pi^0$ from the Belle experiment, $\mathcal{B}(\Lambda_{c}^{+}\to p\pi^0)<8.0 \times 10^{-5}$ at the 90\% confidence level~\cite{Belle}, the ratio of branching fractions between $\Lambda_{c}^{+}\to n \pi^{+}$ and $\Lambda_{c}^{+}\to p\pi^0$ is calculated to be larger than 7.2 at the 90\% confidence level, which disagrees with most predictions of the phenomenological models~\cite{SU(3)1,SU(3)2,Geng2018,Uppal1994,Cheng2018,Geng2019,Zhao2020}.

\subsection{The results in the future}
There are a lot of new results will be published using the data samples from 4.600 to 4.946 GeV. We list some of them in the TABLE~\ref{FutRes}. Most of them use the DT Method and the statistical precision can improve a lot from MC simulation. More new decay modes of $\Lambda_{c}^{+}$ will also be searched using these data samples and will be published very soon. With the largest data samples near $\Lambda_{c}^{+}\bar{\Lambda}_{c}^{-}$ threshold in BESIII, the knowledge about $\Lambda_{c}^{+}$ decays can improve a lot. There is an interest in the $\Lambda_{c}^{+}$ decay and have a more detailed explanation of studying the properties of $\Lambda_{c}^{+}$ decays in~\cite{WhitePaper}.
\begin{table}[h]
	\begin{center}
		\caption{The results in the future.}
		\begin{tabular}{|l|c|c|}
			\hline 
			\textbf{Decay} & \textbf{Method} & \textbf{\makecell[c]{Advantage from \\ MC Simulation}} \\
			\hline
			\multicolumn{3}{|c|}{Hadronic Decays}\\
			\hline
			$\Lambda_{c}^{+}\to p \pi^{0}$ & DT Method & about $3\sigma$ significance\\
			\hline
			$\Lambda_{c}^{+}\to p \eta$ & ST Method & $\sigma_{\mathrm{stat.}}$: 22.6\% $\to$ 8.1\%\\
			\hline
			$\Lambda_{c}^{+}\to p \eta'$ & DT Method & about $4\sigma$ significance\\
			\hline
			\multicolumn{3}{|c|}{Semileptonic Decays}\\
			\hline
			$\Lambda_{c}^{+}\to pK^{-}e^{+}\nu_{e}$ & DT Method & $>5\sigma$ significance\\
			\hline
			$\Lambda_{c}^{+}\to \Lambda e^{+}\nu_{e}$ & DT Method & $\sigma_{\mathrm{stat.}}$: 25\% $\to$ 12\%\\
			\hline
			\multicolumn{3}{|c|}{Inclusive Decays}\\
			\hline
			$\bar{\Lambda}_{c}^{-}\to \bar{n}+X$ & DT Method &  $\sigma_{\mathrm{stat.}}$: 32\% $\to$ 3\%\\
			\hline
			$\Lambda_{c}^{+}\to X+ e^{+}\nu_{e}$ & DT Method & $\sigma_{\mathrm{stat.}}$: 9\% $\to$ 3\%\\

			\hline
		\end{tabular}
		\label{FutRes}
	\end{center}
\end{table}

\section{Summary}
BESIII has made great progress in taking data from 4.612 GeV to 4.946 GeV which are the large data samples near the $\Lambda_{c}^{+}\bar{\Lambda}_{c}^{-}$ threshold~\cite{Lum}. We have measured the branching fraction of $\Lambda_{c}^{+}\to n\pi^{+}$, $\Lambda_{c}^{+}\to\Lambda\pi^{+}$, and $\Lambda_{c}^{+}\to\Sigma^{0}\pi^{+}$ to be $(6.6\pm1.2_{\mathrm{stat}}\pm0.4_{\mathrm{syst}})\times10^{-4}$, $(1.31\pm0.08_{\mathrm{stat}}\pm0.05_{\mathrm{syst}})\times10^{-2}$, and $(1.22\pm0.08_{\mathrm{stat}}\pm0.07_{\mathrm{syst}})\times10^{-2}$, respectively by using data samples corresponding to a total integrated luminosity of 3.9 $\mathrm{fb}^{-1}$ collected between 4.612 and 4.699 GeV with the BESIII detector. Taking the upper limit of the branching fraction of $\Lambda_{c}^{+}\to p\pi^{+}$ from the Belle experiment, $\mathcal{B}(\Lambda_{c}^{+}\to p\pi^{0})<8.0\times10^{-5}$ at the 90\% confidence level, the ratio of branching fractions between $\Lambda_{c}^{+}\to n\pi^{+}$ and $\Lambda_{c}^{+}\to p\pi^{0}$ is calculated to be larger than 7.2 at the 90\% confidence level. In order to obtain an improved understanding it is desirable to perform improved studies of $\Lambda_{c}^{+}\to p\pi^{0}$ branching fraction~\cite{npi}. More studies about the $\Lambda_{c}^{+}$ decays are proceeding and the results will be published very soon.

\begin{acknowledgments}
We thank our BEPCII colleagues for the excellent luminosity and our BESIII collaborators for their many efforts in advancing the knowledge of $\Lambda_{c}^{+}$ decays. We are grateful to the FPCP2022 committee for the organization of this nice conference.
\end{acknowledgments}

\bigskip 

\end{document}